\documentclass[12pt,a4paper]{article}
\usepackage{amsmath}
\usepackage{amsxtra}
    \usepackage{amstext}
    \usepackage{amssymb}
    \usepackage{latexsym}
    \usepackage{graphicx}
\usepackage{color}
\usepackage{graphics}

\newcommand{\hepth}[1]{{\tt hep-th/#1}}

\newcommand{\nn}{\nonumber}

\newcommand{\p}{\vspace{6pt}\noindent}
\newcommand{\jump}{\vspace{2pt}}

\advance\textheight by \topskip
\makeatletter

\@addtoreset{equation}{section}
\def\section{\@startsection {section}{1}{\z@}{-8.5ex plus -1ex minus
 -.2ex}{3.3ex plus .2ex}{\large\bf}}
\def\subsection{\@startsection{subsection}{2}{\z@}{-3.25ex plus
 -1ex minus -.2ex}{1.5ex plus .2ex}{\bf}}
\def\subsubsection{\@startsection{subsubsection}{3}{\z@}{-3.25ex plus%
 -1ex minus -.2ex}{1.5ex plus .2ex}{\sl}}

\begin{document}

\begin{titlepage}
\vspace*{-2cm}
\begin{flushright}
\end{flushright}

\vspace{0.3cm}

\begin{center}
{\Large {\bf }} \vspace{1cm} {\Large {\bf Comments on defects
in the $a_r$  Toda field theories}}\\
\vspace{1cm} {\large  E.\ Corrigan\footnote{\noindent E-mail: {\tt
edward.corrigan@durham.ac.uk}} and
C.\ Zambon\footnote{\noindent E-mail: {\tt cristina.zambon@durham.ac.uk}} \\
\vspace{0.3cm}
{\em Department of Mathematical Sciences \\ University of Durham, Durham DH1 3LE, U.K.}} \\

\vspace{2cm} {\bf{ABSTRACT}}
\end{center}
A simple, basic, argument is given, based solely on energy-momentum considerations
to recover  conditions under which $a_r$ affine or conformal Toda field theories can support
defects of integrable type. Associated triangle relations are solved to provide
expressions for transmission matrices that generalize  previously known examples
calculated for the sine-Gordon model and the $a_2$ affine Toda model.

\vfill
\end{titlepage}

\section{Introduction}

The study of defects, or impurities, within integrable field theory was initiated nearly
fifteen years ago by Delfino, Mussardo and Simonetti \cite{Delf94}. They pointed out
that, with some natural assumptions, it would not be possible for an integrable system 
to encompass a defect, such as a delta-function impurity, allowing both reflection and 
transmission compatible with a non-trivial bulk scattering matrix. One may question the 
assumptions (see, for example \cite{Mintchev02}), or analyse those types of defects that
are compatible with the bulk $S$-matrix, for example, those that are purely transmitting
(within the sine-Gordon model this began with some work of Konik and LeClair 
\cite{Konik97}).
It is already known from numerical studies of phenomena in the classical sine-Gordon 
model that a delta-function impurity is unlikely to be integrable (see, for example 
\cite{Goodman2002}), but it was pointed out in \cite{bczlandau} that another possibility
was to allow field discontinuities. At first sight, this appears to be quite drastic
and unlikely to lead anywhere. However, it turned out that discontinuities could be 
permitted provided the fields on either side of the discontinuity were `sewn' together
appropriately. Moreover, the sewing conditions that guaranteed integrability were
closely related to B\"acklund transformations, a fact that emerged not only in 
the sine-Gordon model but also for the subset of affine Toda models defined in terms
of the root data of $a_r$ (the sine-Gordon model itself corresponding to $a_1$) 
\cite{Bowcock04}. For the sine-Gordon model itself
it was possible to analyse in \cite{bczsg} the relationship between the classical and 
quantum theories 
possessing this type of discontinuity - which are really more akin to `shocks',
 and sometimes
referred to as `jump-defects' to distinguish them from delta-function impurities - thereby
providing a framework for the Konik-LeClair transmission matrix and various means of 
checking it, including perturbative calculations of the transmission factors for breathers
\cite{Bajnok07}.

\p One purpose of this article is to provide  simple and reasonably general arguments 
leading to the sewing conditions previously proposed for the affine Toda field theories.
It appears the $a_r$ models are special and we have not yet found a way to generalize the
argument to all the other models, or indeed to find an alternative. This is a slightly
frustrating situation, perhaps indicating simply a lack of imagination, because in
other contexts members of the whole class of affine Toda models, apart from relatively
small details depening on the choice of root system, have very similar properties.
In passing, it is remarked how in the context of the conformal Toda models a sequence of
defects can transform one model into another. The illustrative example of this behaviour
shows how an $a_r$ model can be reduced to an $a_{r-1}$ model together with a free
massless field. Once this is shown to be integrable (and an argument is provided in 
section 3), combinations of defects can be used to construct mixtures of conformal
models. The simplest example of this is the well-known relationship between the 
Liouville model and a massless free field.

\p A second purpose is to make progress towards completing
 the story that was begun in \cite{Corrigan07}. There, besides general remarks
that applied to each member of the $a_r$ class of affine Toda models, 
it proved possible to
solve in detail the triangle relations for $a_2$ affine Toda theory and to 
describe some of the
intriguing properties of the transmission matrix, especially those surrounding the 
curiously different character of the interactions between the defect and the two
types of soliton (perhaps better regarded as soliton and anti-soliton). Here,
the techniques are generalized to calculate  transmission matrices for the $a_r$
affine Toda models and to investigate some of their properties, particularly
with regard to unstable bound states.

\section{Energy and momentum revisited}
\label{MomentumConservation}

\p In the bulk, $-\infty <x<\infty$, an affine Toda field theory
corresponding to the root data of a Lie algebra $g$ is
described  by the Lagrangian density
\begin{equation}\label{TodaL}
\mathcal{L}_u =\frac{1}{2}\left( \partial_{\mu}u\cdot
\partial^{\mu}u\right)-\frac{m^{2}}
 {{\beta}^{2}}\sum_{j=0}^{r}\, n_j\,(e^{{\beta}\alpha_{j}\cdot u}-1),
\end{equation}
where $m$ and ${\beta}$ are constants and $r$ is the rank of the
algebra. The set of vectors $\{\alpha_j\}$ with $j=1,\dots,r$ are
the simple roots of $g$, while
$\alpha_0$ is an extra  root, defined by
$\alpha_0=-\sum_{j=1}^{r}\,n_j\,\alpha_j.$
The integers $\{n_j\}$ are a set of integers characteristic of each affine
Toda model.
Each set of roots is associated with an affine
Dynkin-Ka\v c diagram, which encodes the inner products among the simple
roots $\{\alpha_j\}$ including the extra root $\alpha_0$.
Finally, the field $u=(u_1,u_2,\, \dots,\, u_r)$ takes values in the
$r$-dimensional Euclidean space spanned by the simple roots $\{\alpha_j\}$.
The affine Toda models are
massive and integrable. However, if the term in the Lagrangian with $j=0$
is omitted, then the theory described by the density Lagrangian \eqref{TodaL}
is conformal and called a conformal Toda field theory.
All these models posses a Lax pair representation and they have been
extensively investigated in the past,
both classically and in the quantum domain.
For further details concerning the affine Toda
field theories, see \cite{Arinshtein79, Braden90}, and the review
\cite{Corrigan94};  for further details on the conformal Toda models
see, for instance, \cite{Mansfield83,Fateev88}, and references therein.

\p In the present article, a Lagrangian density of the following type
\begin{equation}\label{DefectLagrangian}
{\cal L}_{D}=\theta(-x){\cal L}_u +\theta(x){\cal
L}_v-\delta(x)\mathcal{W},
\end{equation}
which couples together two sets of $r$ scalar fields $u$, $v$
by means of a defect located in $x=0$, will be investigated.

\p The purpose of this section is to start from first principles to determine
for which Toda field theories there is a set of defect conditions that will
allow exchange of energy-momentum between a defect and the fields
on either side of it. The result is a little surprising.

\p Consider the standard expressions for the energy and momentum carried
by the fields $u$ and $v$:
$$
{\cal E}=\int_{-\infty}^0 dx\,\left(\frac{1}{2}(u_x\cdot u_x)+
\frac{1}{2}(u_t\cdot u_t)+U(u)\right)+\int_0^{\infty}dx \,
\left(\frac{1}{2}(u_x\cdot u_x)+\frac{1}{2}(v_t\cdot v_t)+V(v)\right),
$$
and
$$
{\cal P}=\int_{-\infty}^0 dx\,(u_x\cdot u_t)+\int_0^{\infty}dx \,(v_x\cdot v_t),
$$
where, for the time being, the potentials for the fields $u$ and $v$
remain unspecified.
Differentiating with respect to time, using the equations of motion
for the two fields in their respective domains, and assuming no
contributions from $x=\pm\infty$ give, one has, respectively,
\begin{equation}\label{EnergyTimeDerivative}
\dot{\cal E}=\left. u_x\cdot u_t\right|_{x=0}-\left. v_x\cdot v_t\right|_{x=0},
\end{equation}
and
\begin{equation}\label{MomentumTimeDerivative}
\dot{\cal P}=\left(\frac{1}{2}(u_x\cdot u_x)+\frac{1}{2}
(u_t\cdot u_t)-U(u)\right)_{x=0}-\left(\frac{1}{2}(v_x\cdot v_x)+
\frac{1}{2}(v_t\cdot v_t)-V(v)\right)_{x=0}.
\end{equation}
Energy-momentum will be exchangeable with the defect provided each
of these may be
expressed as time derivatives of  functions of the fields evaluated at $x=0$.
Consider first \eqref{EnergyTimeDerivative} and suppose, at $x=0$, the rather
general condition relating space derivatives,
\begin{equation}\label{DefectConditions}
u_x=Au_t +Bv_t + X(u,v),\quad v_x=Cu_t +Dv_t +Y(u,v),
\end{equation}
where $A,B,C,D$ are  matrices, and $X,Y$ are vector functions of
$u$ and $v$. Then,
$$\dot{\cal E}=u_t\cdot Au_t +u_t\cdot Bv_t-v_t\cdot Cu_t -
v_t\cdot Dv_t +u_t\cdot X
-v_t\cdot Y,
$$
and this will be a total time derivative provided
\begin{equation}
C=B^T, \quad A=-A^T, \quad D=-D^T, \quad X=-\nabla_u {\cal D},
\quad Y=\nabla_v{\cal D},
\end{equation}
where ${\cal D}$ is also a function of $u$ and $v$ evaluated at
$x=0$. Under these circumstances, ${\cal E+D}$ is conserved.

\p Next, consider \eqref{MomentumTimeDerivative}. A similar
computation places further constraints, namely
\begin{equation}\label{MomentumConservationConstraints1set}
1-A^2=BB^T,\quad 1-D^2=B^TB,\quad AB+BD=0,
\end{equation}
and
\begin{equation}\label{MomentumConservationConstraints3set}
(\nabla_u {\cal D}\cdot \nabla_u {\cal D})-(\nabla_v{\cal D}
\cdot \nabla_v{\cal D})=2(U-V),
\end{equation}
together with the requirement
\begin{equation}\label{MomentumConservationConstraints2set}
(A\nabla_u {\cal D}-B\nabla_v{\cal D})=\nabla_u\Omega,\quad
(B^T\nabla_u {\cal D}-D\nabla_v{\cal D})=-\nabla_v\Omega,
\end{equation}
where $\Omega$ is another function of the fields $u$ and $v$ evaluated at $x=0$.
Under these circumstances, ${\cal P}+\Omega$ is conserved.
Provided all these constraints may be simultaneously satisfied energy
and momentum will be conserved once specific contributions coming from
the defect itself are taken into account.

\p The first expression in \eqref{MomentumConservationConstraints1set}
may be rewritten as follows
\begin{equation}\label{MKMatrixRelation}
(1+A^T)(1+A)=BB^T.
\end{equation}
Since $A$ is real and antisymmetric, its eigenvalues are purely
imaginary and hence $(1-A)$ is invertible. Then
\begin{equation}
1=(1-A)^{-1}B((1-A)^{-1}B)^T,
\end{equation}
which implies that $(1-A)^{-1}B$ is orthogonal and $B=(1-A)\,O$,
with $O\in O(r)$.

\p At this stage, if it is further assumed that the various matrices
are independent of $u$ and $v$, it can be remarked also that the
set of boundary conditions of the type \eqref{DefectConditions}
follows from the Lagrangian density \eqref{DefectLagrangian} with the choice
\begin{equation}\label{DefectLagrangian}
\mathcal{W}=\left(\frac{1}{2}u_t\cdot
A u -\frac{1}{2}v_t \cdot D v+u_t\cdot B v+\mathcal{D}(u, v )\right).
\end{equation}
Making an orthogonal transformation on the field $v$,
this expression may be rewritten
\begin{equation}
\mathcal{W}=\left(\frac{1}{2}u_t\cdot
A u -\frac{1}{2}v_t \cdot D' v+u_t \cdot(1-A) v+\mathcal{D}(u, v )\right).
\end{equation}
As a consequence, the third expression in
\eqref{MomentumConservationConstraints1set} reads
\begin{equation}\label{D'}
A(1-A)+(1-A)D'=0,
\end{equation}
which implies $D'=-A$ and
\begin{equation}\label{B}
B=(1-A), \quad B^T=(1+A)=(2-B).
\end{equation}
It is easy to verify that the second equation in
\eqref{MomentumConservationConstraints1set} is automatically satisfied.
It is worth pointing out that an orthogonal transformation on the
field $v$ translates simply into a change of base for the simple roots
appearing in the expression of the potential $V(v)$.
Note that a similar result would have been  obtained by starting from the
second equation in \eqref{MomentumConservationConstraints1set} and
applying an orthogonal transformation to the field $u$.

\p Results \eqref{D'}, \eqref{B} allow
constraints \eqref{MomentumConservationConstraints2set}
to be rewritten in terms of the matrix $A$ alone. Then, by computing
all the second derivatives of $\Omega$
 and requiring consistency, the following
additional constraints are obtained
\begin{eqnarray}\label{CrossDerivatives}
A_{lj}\,\frac{\partial^2 \mathcal{D}}{\partial u_k \,
\partial u_j}-(1-A)_{ij}\,
\frac{\partial \mathcal{D}}{\partial u_k\, \partial v_j}&=&A_{kj}\,
\frac{\partial^2 \mathcal{D}}{\partial u_l \,\partial u_j}-(1-A)_{kj}\,
\frac{\partial \mathcal{D}}{\partial u_l\, \partial v_j},\nn\\
(1+A)_{lj}\,
\frac{\partial \mathcal{D}}{\partial v_k\, \partial u_j}+A_{lj}\,
\frac{\partial^2 \mathcal{D}}{\partial v_k \,\partial v_j}&=&(1+A)_{kj}\,
\frac{\partial \mathcal{D}}{\partial v_l\, \partial u_j}+A_{kj}\,
\frac{\partial^2 \mathcal{D}}{\partial v_l \,\partial v_j},\nn\\
-A_{lj}\,\frac{\partial^2 \mathcal{D}}{\partial v_k \,\partial u_j}+(1-A)_{ij}\,
\frac{\partial \mathcal{D}}{\partial v_k\, \partial v_j}&=&(1+A)_{kj}\,
\frac{\partial \mathcal{D}}{\partial u_l\, \partial u_j}+A_{kj}\,
\frac{\partial^2 \mathcal{D}}{\partial u_l \,\partial v_j}.
\end{eqnarray}

\p Since $u$, $v$ are  Toda-like fields,
solutions for the defect potential $\mathcal{D}$ should have  the form
$$\exp{(a\cdot u+b\cdot v)}$$ where $a$, $b$ are vectors needing to be specified.
Using this fact the constraints \eqref{CrossDerivatives} reduce
to the following tensorial expressions
\begin{eqnarray}
a\otimes (-A\,a+(1-A)\,b)&=&(-A\,a+(1-A)\,b)\otimes a \label{CrossDerivative1}\\
((1+A)\,a+A\,b)\otimes b &=&b\otimes ((1+A)\,a+A\,b)\label{CrossDerivative2}\\
b\otimes(-A\,a+(1-A)\,b)&=&((1+A)\,a+A\,b)\otimes a.\label{CrossDerivative3}
\end{eqnarray}
The first two relations imply $(-A\,a+(1-A)\,b)=\alpha\,a$, and
$((1+A)\,a+A\,b)=\beta\, b$,
respectively, with $\alpha$, $\beta$ constants. Hence \eqref{CrossDerivative3}
forces $\alpha=\beta$.
Constraints for the vectors $a$ and $b$ are provided by \eqref{CrossDerivative1}
and \eqref{CrossDerivative2}, since they may be rewritten as follows
\begin{equation}
a=(1+A)^{-1}\,(\alpha-A)\,b,\qquad (1-\alpha^2)\,b=0.
\end{equation}
Clearly $\alpha^2=1$ since the possibility $b=0$ is uninteresting since it
also implies $a=0$ and a trivial ${\cal D}$.
Choosing $\alpha=1$ and setting $a=(1-A)\,x/2$, $b=(1+A)\,x/2$ the defect
potential $\mathcal{D}$ has the form $\exp{((u\cdot(1-A)+v\cdot(1+A)) x/2)}$,
while
for $\alpha=-1$, setting $a=-b=y/2$, the defect potential $\mathcal{D}$
has the form $\exp{((u-v)\cdot y/2)}$. The vectors $x$, $y$ are not
yet determined.

\p Information collected so far implies a general expression
for the defect potential, namely,
\begin{equation}\label{DefectPotential}
\mathcal{D}=\sum_k\, p_k\, e^{(u-v)\cdot y_k/2}+ \sum_l\, q_l\,
e^{(u\cdot(1-A)+v\cdot(1+A)) x_l/2},
\end{equation}
where $p_k$, $q_l$ are constant coefficients.

\p This expression can now
be used to investigate the last constraint
\eqref{MomentumConservationConstraints3set}, which links the defect potential to
the bulk potentials for the fields $u$ and $v$.
After some algebra the constraint turns out to be:
\begin{equation}\label{LastConstraint}
\sum_{k,l}\, p_k\,q_l\, (y_k\cdot x_l) \,
e^{u\cdot(y_k+(1-A)x_l)/2+v\cdot(-y_k+(1+A)x_l)/2}=2(U(u)-V(v)).
\end{equation}
Before  analyzing this expression, note that on the left
hand side of \eqref{LastConstraint} there can be no repeated exponents.
In fact, if two exponents were to be the same, given two pairs
of vectors ($x_i, y_i$) and ($x_j, y_j$), the following condition  would hold
$$
y_i+(1-A)\,x_i=y_j+(1-A)\,x_j, \quad -y_i+(1+A)\,x_i=-y_j+(1+A)\,x_j,
$$
implying $x_i=x_j$ and, therefore, $y_i=y_j$.
Hence, to each different pair of vectors
($x_i, y_i$) and ($x_j, y_j$) there correspond two different exponents
in the left
hand side of expression \eqref{LastConstraint}. Note also that,
in principle, the two bulk potentials $U(u)$ and $V(v)$ could belong to two
different Toda-like models, provided  the number $r$ of fields either side
of the defect is the same.

\p Denote by $\{\alpha_i\}$, $\{\alpha_i'\}$ the two sets of simple roots
of the Lie algebras associated with the models on the left and on the right
of the defect respectively, together with - if required -
the extended root.
Since the left hand side of \eqref{LastConstraint} must be equal to the
difference of two Toda-like
bulk potentials, there must exist four sets of vectors $\{x_l\}$,
$\{x_l'\}$, $\{y_k\}$, $\{y_k'\}$ such that
\begin{equation}\label{potentialu}
y_i=(1+A)\,x_i, \quad (y_i\cdot x_i)\neq 0, \quad y_i\in\{y_k\},\,x_i\in
\{x_l\}\;\ \hbox{to\ give}\ \  e^{u\cdot x_i},
\end{equation}
and
\begin{equation}\label{potentialv}
y_i'=-(1-A)\,x_i', \quad (y_i'\cdot x_i')\neq 0, \quad y_i'\in\{y_k'\},\,
x_i'\in \{x_l'\}\; \ \hbox{to\ give}\ \  e^{v\cdot x_i'}.
\end{equation}
Clearly,
$x_i\equiv\alpha_i$, $x_i'\equiv\alpha_i'$.
The exponential terms obtained in \eqref{potentialu} and \eqref{potentialv}
correspond to pieces necessary for building the two bulk potentials $U(u)$
and $V(v)$. Obviously, they are the only possibilities allowed in
\eqref{LastConstraint}.
This means that any other terms that might arise  in \eqref{LastConstraint}
because
of particular choices of vectors $\{\alpha_l\}$, $\{\alpha_l'\}$, $\{y_k\}$,
$\{y_k'\}$
must have  coefficients equal to zero. Note, the possibility of having two
exponentials that
cancel is ruled out by the fact that any given exponential may only appear once
on the left hand side of \eqref{LastConstraint}, as established above.

\p To analyze further expression \eqref{LastConstraint}, consider first the
case
in which the two sets $\{\alpha_l\}$ and  $\{\alpha_l'\}$ coincide.
For a given
$y_i\in\{y_k\}$, it could happen that
\begin{equation}\label{condition1}
(y_i\cdot \alpha_j)=0 \quad \forall \alpha_j\in\{\alpha_l\}\; j\neq i.
\end{equation}
Hence, it would be possible to write $y_i=(\alpha_i \cdot \alpha_i)\, w_i$,
where $w_i$
is a fundamental highest weight of the Lie algebra associated with the
Toda-like model on both sides of the defect\footnote{In the present case they coincide
since they have the same set of simple roots.}.
Note that this choice satisfies condition \eqref{condition1} and, because of
\eqref{potentialu}, implies
$\alpha_j\cdot A \alpha_i=-\alpha_j\cdot \alpha_i, \; \forall j\neq i$.

\p As an alternative to condition \eqref{condition1}, suppose that for a
$y_i\in\{y_k\}$ there exists at least one $\alpha_j\in\{\alpha_l\}$ such
that $(y_i\cdot \alpha_j)\neq 0$ with $j\neq i$. In the most general
case, the exponent associated with the pair ($y_i, \alpha_j$) in
\eqref{LastConstraint} will be a combination of both fields $u$ and $v$.
However, such a term  is not allowed. A way out is to suppose, in addition,
 $y_i=-(1-A)\,\alpha_j$.
Then, $y_i\equiv y_j'$ with $y_j'\in\{y_k'\}$, therefore the resulting
exponential
is permitted since it coincides with a term of the bulk potential $V(v)$. Then,
\begin{equation}
y_i=2\alpha_i-(1-A)\,\alpha_i, \qquad y_i=-2\alpha_j+(1+A)\,\alpha_j.
\end{equation}
Multiplying these two expressions by $\alpha_i$ and $\alpha_j$,
respectively,  leads to
\begin{equation}
(\alpha_i \cdot \alpha_i)=(\alpha_j \cdot \alpha_j)=
(y_i \cdot\alpha_i)=-(y_i \cdot \alpha_j).
\end{equation}
Hence it is possible to write $y_i=(\alpha_i \cdot \alpha_i)(w_i-w_j)$
where $w_i$, $w_j$ are fundamental highest weights of the Lie algebra associated
with the two Toda-like models. Note, this situation can only occur
when the roots $\alpha_i$, $\alpha_j$ have the same length.

\p In summary, the expression \eqref{LastConstraint} is solved by choosing
three sets $\{x_l\}$,
$\{y_k\}$ and $\{y_k'\}$ such that
$$\{x_l\}\equiv \{\alpha_l\},$$
where $\{\alpha_l\}$ is a set of simple roots together with - if included -
the extended root, and $y_i\in\{y_k\}$, $y_i'\in\{y_k'\}$ can have one of the
following form ($(a)$, $(c)$ for $y_i$ and $(b)$, $(c)$ for $y_i'$,
respectively):
\begin{eqnarray}
(a)\quad & y_i=(1+A)\,\alpha_i=(\alpha_i \cdot \alpha_i)\,w_i,
\quad (\alpha_j \cdot A\,\alpha_i)=-(\alpha_j\cdot \alpha_i)
\quad j\neq i \label{possibilitya}\\
(b)\quad & y_i'=-(1-A)\,\alpha_i=-(\alpha_i \cdot \alpha_i)\,w_i,
\quad (\alpha_j \cdot A\,\alpha_i)=(\alpha_j\cdot \alpha_i) \quad j\neq i
\label{possibilityb}\\
(c)\quad & y_i\equiv y_j'=(1+A)\,\alpha_i=-(1-A)\,\alpha_j=(\alpha_i \cdot
\alpha_i)\,(w_i-w_j), \nonumber \\
& (\alpha_j \cdot A\,\alpha_i)=-(\alpha_i\cdot \alpha_i)-(\alpha_j\cdot \alpha_i),\quad
(\alpha_i\cdot\alpha_i)\equiv(\alpha_j\cdot \alpha_j)\quad j\neq i. \label{possibilityc}
\end{eqnarray}
Clearly, the possibility $(c)$ \eqref{possibilityc} implies an overlapping
among the elements in the two sets $\{y_k\}$ and $\{y_k'\}$.
To decide which form among the possibilities listed above to choose for
each vector in these two sets, it is worth noticing that because of
\eqref{potentialu},
\eqref{potentialv},
the following constraint holds
\begin{equation}
y_i-y_i'=2\,\alpha_i.
\end{equation}
To satisfy this constraint, the only possible combinations for the
explicit forms of the pair $(y_i,y_i')$ are
\begin{eqnarray}
((a), (b))\;&  y_i-y_i'=2\,\alpha_i=2\,(\alpha_i \cdot \alpha_i)\,w_i
\label{combinationab} \\
((a), (c))\; & y_i-y_i'=2\,\alpha_i=(\alpha_i \cdot \alpha_i)\,(2\,w_i-w_j),
\ (1+A)\,\alpha_j=-(1-A)\,\alpha_i,\;\; i\neq j \label{combinationac} \\
((c), (b))\;&  y_i-y_i'=2\,\alpha_i=(\alpha_i \cdot \alpha_i)\,(2\,w_i-w_m),
\ (1+A)\,\alpha_i=-(1-A)\,\alpha_m,\;\; i\neq m \label{combinationcb} \\
((c), (c))\;&  y_i-y_i'=2\,\alpha_i=(\alpha_i \cdot \alpha_i)\,(2\,w_i-w_j-w_m),
\label{combinationcc} \nn\\
 & (1+A)\,\alpha_i=-(1-A)\,\alpha_j,
\;(1+A)\,\alpha_m=-(1-A)\,\alpha_i,\;\; i\neq j\neq m.
\end{eqnarray}
Clearly the combination \eqref{combinationab} can only appear if the root is
unconnected to all the others since there
is no simple root that coincides up to scaling with a single
fundamental highest weight except for $a_1$. Such cases will not be considered further
here since the Dynkin diagram would have at least one disconnected spot.
In addition, by looking at the other combinations, it is clear
that each node of the Dynkin diagram associated to the set of roots
$\{\alpha_l\}$
must have no more than two linked neighbours
since $\alpha_i=(2\,w_i-w_m)$ or $\alpha_i=(2\,w_i-w_j-w_m)$
at most\footnote{Setting $(\alpha_i \cdot \alpha_i)=2$}. This fact, together
with the observation that the possibility $(c)$ \eqref{possibilityc} might
only happen when the roots involved have the same length, implies that the
only Toda-like field theories allowed are those associated to  Lie algebras
of  type $a_r$. This is a surprising result and the only assumption made
 to simplify the discussion was to suppose the matrices $A,B,C,D$ were
independent of $u$ and $v$. Relaxing this considerably complicates the
discussion yet might be necessary to be able to apply the same type
of arguments to Toda models based on other root systems.

\p From now on consider only the Lie algebra $a_r$. The simple roots
together with
the extended (lowest) root can be written in term of the fundamental
highest weights  $w_i$ $i=1,\dots,r$ via
\begin{equation}
\alpha_i=(2\,w_i-w_{i+1}-w_{i-1})\quad i=1,\dots,r,\qquad w_0\equiv w_{r+1}=0.
\end{equation}
Consider two affine $a_r$ Toda theories on either side of the defect, then
a total momentum is conserved provided
$\{\alpha_l\}$ is the set of simple roots of $a_r$ together with the
extended root, and the elements of the sets $\{y_k\}$, $\{y_k'\}$
with $k,k'= 0,\dots,r$
are as follows
\begin{equation}\label{Bmatrix0y}
y_i=y_{i-1}'\quad i=2,\dots,r,\qquad y_1=y_0',\qquad y_0=y_r'.
\end{equation}
It can be noticed that the pairs $(y_1,y_1')$, $(y_r,y_r')$ correspond
to the combinations \eqref{combinationac} and \eqref{combinationcb},
respectively, while
all other pairs correspond to the case \eqref{combinationcc}.
The matrix $B$ may be written explicitly in terms of the fundamental
weights of the algebra $a_r$
\begin{equation}\label{Bmatrix0}
B=(1-A)=2\sum_{a=1}^{r}\,(w_a-w_{a+1})\,w_a^T,
\end{equation}
A formula first obtained by other means in \cite{Bowcock04}.

\p Notice that the matrix $B$ can be replaced by its transpose to provide
another solution, which is represented by
\begin{equation}\label{Bmatrix}
B=(1-A)=2\sum_{a=1}^{r}\,w_a\,(w_a-w_{a+1})^T.
\end{equation}
Then
\begin{equation}\label{Bmatrixy}
y_i=y_{i+1}'\quad i=1,\dots,r,\qquad y_0=y_1',
\end{equation}
and the pairs $(y_1,y_1')$, $(y_r,y_r')$ correspond to the combinations
\eqref{combinationcb} and \eqref{combinationac}, respectively, while
all other pairs correspond to the case \eqref{combinationcc}.

\p Setting $p_i=\sigma$ for all $p_i\in\{p_k\}$ and $q_i=1/\sigma$ for all
$q_i\in\{q_l\}$, the defect potential \eqref{DefectPotential} reads
\begin{equation}\label{DefectPotential0}
\mathcal{D}=\sum_{k=0}^{r}\, \frac{1}{\sigma}\, e^{(u-v)\cdot (1+A)\,
\alpha_k/2}+ \sum_{l=0}^r\,\sigma\, e^{(u\cdot(1-A)+v\cdot(1+A))\, \alpha_l/2}.
\end{equation}

\p Next, consider instead two $a_r$ conformal Toda theories on either
side of the
defect. A solution to expression \eqref{LastConstraint} is represented,
for example, by the  matrix $B$ as in \eqref{Bmatrix0} with
$\{\alpha_l\}$ the set of simple roots of $a_r$ and the
two sets $\{y_k\}$, $\{y_k'\}$ as in \eqref{Bmatrix0y} with $k,k'= 1,\dots,r$.
In other words,  the vectors $y_0$ and $y_0'$ have been omitted.
The defect potential is
\begin{equation}
\mathcal{D}=\sum_{k=1}^{r}\, \frac{1}{\sigma}\, e^{(u-v)\cdot (1+A)\,\alpha_k/2}+
\frac{1}{\sigma}\, e^{-(u-v)\cdot (1-A)\,\alpha_r/2} + \sum_{l=1}^r\, \sigma\,
e^{(u\cdot(1-A)+v\cdot(1+A))\, \alpha_l/2}.
\end{equation}
But, from \eqref{Bmatrix0}, it is easy to realize that $-(1-A)\,\alpha_r=(1+A)\,
\alpha_0$, where $\alpha_0$ is the extended root, hence
\begin{equation}\label{DefectPotentialConformal}
\mathcal{D}=\sum_{k=0}^{r}\, \frac{1}{\sigma}\, e^{(u-v)\cdot (1+A)\,\alpha_k/2}+
\sum_{l=1}^r\, \sigma\, e^{(u\cdot(1-A)+v\cdot(1+A))\, \alpha_l/2}.
\end{equation}

\p
Finally, there is another intriguing possibility. Consider, for example,
the matrix \eqref{Bmatrix0}. Then a solution of expression \eqref{LastConstraint}
 is provided by a set $\{\alpha_l\}$ of simple roots of the Lie algebra $a_r$ and
 two sets $\{y_k\}$ k= 1,\dots,r and $\{y_k'\}$ with $k'= 1,\dots,(r-1)$, whose
elements satisfy \eqref{Bmatrix0y}. This time the vectors
$y_0$, $y_0'$ and $y_r'$ are
missing. Under these circumstances, the defect potential read
\begin{equation}\label{DefectPotentialConformal+free}
\mathcal{D}=\sum_{k=1}^{r}\, \frac{1}{\sigma}\, e^{(u-v)\cdot (1+A)\,\alpha_k/2}+
\sum_{l=1}^r\, \sigma\, e^{(u\cdot(1-A)+v\cdot(1+A))\, \alpha_l/2}.
\end{equation}
This situation allows the conservation of momentum for a defect system with an
$a_r$ conformal Toda field theory on the left of the defect and an $a_{r-1}$
conformal Toda theory plus a free massless field, on the right. This case is
allowed since all the fields involved are massless.  In
fact, it is possible to think of the algebra $a_{r-1}$ embedded within the
$a_{r}$ algebra and the defect peels off the simple root at one end of the Dynkin
diagram. If there was a sequence of $r$ defects it would be possible to reduce
the $a_r$ conformal Toda theory to a collection of free massless fields,
a situation that was not noticed before.

\p To clarify this point and verify that the conservation of
a total momentum implies integrability, the Lax pair construction for this
specific case will be explored in the next section.

\p Before concluding this section, it is worth adding a few words on the
possibility that
the two sets $\{\alpha_l\}$ and  $\{\alpha_l'\}$ do not in fact coincide.
First of all, it is clear these two sets cannot be completely disjoint.
In fact, if this were the case, then, in addition to \eqref{potentialu}
and \eqref{potentialv},  for each $y_i\in \{y_k\}$ it would also be required that
\begin{equation}\label{condition0}
(y_i\cdot \alpha_i')=0, \qquad \forall \alpha_i'\in\{\alpha_k'\}.
\end{equation}
However, the simple roots in $\{\alpha_k'\}$ are $r$ linearly independent
vectors in an $r$ dimensional space, hence condition \eqref{condition0}
could be satisfied only provided $y_i=0$, which is false.

\p Actually, even a partial identification among the elements of the
sets $\{\alpha_l\}$ and  $\{\alpha_l'\}$ is not possible. To see this,
consider two non-orthogonal simple roots $\alpha$ and $\beta$,
$(\alpha\cdot\beta\neq 0)$, such that $y=(1+A)\cdot\alpha$
and $z=(1+A)\cdot\beta$. In this way they will realize
two exponents of the type \eqref{potentialu}, which are part
of the potential $U(u)$. Since the roots are not orthogonal
and the matrix $A$ is antisymmetric, it follows the scalar
products $(\beta\cdot y)$ and $(\alpha\cdot z)$ cannot  both
be zero. Suppose $(\beta\cdot y)\neq 0$. This means that
given the vector $y$ there are two simple roots $\alpha$ and
$\beta$ whose scalar product with $y$ differs from zero.
Then, $y=(1+A)\cdot\alpha$, yet also  $y=-(1-A)\cdot\beta$.
Thus $\beta$ is also a simple root within the set $\{\alpha_l'\}$. In
other words, the simple root $\beta$ is located each of the sets $\{\alpha_l\}$
and  $\{\alpha_l'\}$. Bearing in mind that for each simple root there
is always another simple root that is not orthogonal to it, and continuing
the previous argument for each simple root in either set, it is
inevitable the two sets of simple roots
must coincide. In addition, it emerges that the possible relations
amongst the elements of the sets $\{y_k\}$ and $\{y_k'\}$ - according
to the definition \eqref{potentialu} and \eqref{potentialv} - are
those previously demonstrated for the $a_r$ Lie algebra case (see for
instance \eqref{Bmatrix0y} and \eqref{Bmatrixy}).

\section{The Lax pair construction}

\p In the bulk, a Lax pair representation for a theory with $r$ field
of which $(r-1)$ represents a conformal $a_{r-1}$ Toda theory and the
remaining field a free massless one may have the following form

\begin{eqnarray}\label{TheLaxPair}
a_t&=& \frac{1}{2}\left[\partial_x v \cdot
{\bf H}+ \sum_{i=1}^{r-1}\left(\lambda\,
E_{\alpha_i}-\frac{1}{\lambda}\,
E_{-\alpha_i}\right)e^{\alpha_i\cdot v/{2}}\right]+\lambda \,E_{\alpha_r}\,
e^{\alpha_r\cdot v/2}, \nonumber\\
a_x&=&\frac{1}{2}\left[\partial_t v \cdot
{\bf H}+ \sum_{i=1}^{r-1}\left(\lambda\,
E_{\alpha_i}+\frac{1}{\lambda}\,
E_{-\alpha_i}\right)e^{\alpha_i\cdot v/{2}}\right]+\lambda \,E_{\alpha_r}\,
e^{\alpha_r\cdot v/2}.
\end{eqnarray}
Matrices $\bf{H}$ are the generators of the Cartan subalgebra of a $a_{r}$
Lie algebra whose simples roots are $\alpha_1$, $i=1,\dots,r$ and
$E_{\pm \alpha_i}$ are the generators of the simple roots
or their negatives. Finally, $\lambda$ is the spectral parameter.
Using the Lie algebra commuting relations
\begin{equation}\label{LieAlgebra}
[{\bf{H}},E_{\pm \alpha_i}]=\pm \alpha_i\,E_{\pm \alpha_i},
\qquad [E_{\alpha_i},E_{-\alpha_j}]=\delta_{ij}\,{\bf{H}},
\end{equation}
it can be checked that the Lax pair \eqref{TheLaxPair} ensures
the zero curvature condition
\begin{equation}\label{ZeroCurvatureCondition}
\partial_t a_x-\partial_x a_t+
[a_t,a_x]=0,
\end{equation}
is equivalent to the equation of motion. In the present case,
\begin{equation}\label{EquationOfMotion}
\partial^2 v=-\sum_{i=1}^{r-1}\,\alpha_i\, e^{\alpha_i\cdot v}.
\end{equation}
Thus, the $r$ components of the vector $v$, or $r$
linear combinations of these components - depending on the base chosen
for the simple roots - represent an $a_{r-1}$ Toda field theory together
with a free massless field.

\p Consider a defect at $x=0$, which links an $a_{r}$ Toda field theory
on the left with an $a_{r-1}$ Toda field theory and a free massless
field on the right. The Lax pair describing such a system may be
constructed as explained in \cite{Bowcock04}. Consider two overlapping regions
$R^{<}$ ($x<b,\, b>0$) and $R^{>}$ ($x>a, \, a<0$) each containing
the defect, and in each region define a new Lax pair as follows
\begin{eqnarray}\label{ModifiedLaxPairs}
R^{<}:&&
\hat a_t^{<}=
a_t^{<}(u)-\frac{1}{2}\theta(x-a)\, (u_x -A u_t
- Bv_t \psi+\nabla_u \mathcal{D})\cdot{\bf H},\nn\\
&&\hat a_x^{<}= \theta(a-x)\,a_x^{<}(u),\nn\\
R^{>}:&&
\hat a_t^{>}=
a_t^{>}(v)-\frac{1}{2}\theta(b-x)\, (v_x -B^T u_t
+A v_t -\nabla_v \mathcal{D})\cdot{\bf H},\nn\\
&&\hat a_x^{>}= \theta(x-b)\,a_x^{>}(v),
\end{eqnarray}
where $a_t^{<}$ and $a_x^{<}$ are the Lax pair for an $a_r$ Toda model
(see \cite{Bowcock04}), while $a_t^{>}$ and $a_x^{>}$ coincide with
the Lax pair \eqref{TheLaxPair}.
Applying the zero curvature condition \eqref{ZeroCurvatureCondition},
the Lax pair \eqref{ModifiedLaxPairs} yields both the
equations of motion for the fields $u$ and $v$ in the two regions $x<a$ and
$x>b$ and the defect conditions at $x=0$ and $x=b$. In the overlapping region
$a<x<b$ it implies that the fields $u$ and $v$ are independent of $x$
throughout the overlap. On the other hand, maintaining the
zero curvature condition within the overlap also requires the two components
$\hat a_t^{<}$ and $\hat a_t^{>}$ to be related by a gauge transformation:
\begin{equation}\label{LinkingCondition}
{\cal K}_t={\cal K}\,\hat a_t^{>}- \hat a_t^{<}\,{\cal K}.
\end{equation}
Setting
\begin{equation}\label{GaugeTransformationOnK}
{\cal K}=e^{-{\bf H}\cdot(A u+B v)/2}\,\tilde{\cal K}\,
e^{{\bf H}\cdot(B^{T}u-A v)/2},
\end{equation}
such that the matrix $\tilde{\cal K}$ is independent of $t$, equation
\eqref{LinkingCondition} leads to
\begin{eqnarray}\label{Kequation}
\tilde{\cal K}\, {\bf H}\cdot\nabla_v\mathcal{D}+ {\bf H}\,
\tilde{\cal K}\cdot\nabla_u\mathcal{D}&=& \lambda\sum_{i=1}^r
\,e^{\alpha_i\cdot(B^T u+B v)/2}\,[E_{\alpha_i},\tilde{\cal K}] \nn\\
&&-\frac{1}{\lambda}\sum_{i=1}^r \,E_{-\alpha_i}\,\tilde{\cal
K} \,e^{\alpha_i\cdot B(u-v)/2}+\frac{1}{\lambda}\sum_{i=1}^{r-1}\, \tilde{\cal
K}\,E_{-\alpha_i}\, e^{-\alpha_i\cdot B^T(u-v)/2}.\nn
\end{eqnarray}
Choosing, as an example, the matrix \eqref{Bmatrix0}, the relation
$(\alpha_i\,B^T)=-(\alpha_{i+1}\,B)$ holds, and this allows the above expression
to be rewritten as follows:
\begin{eqnarray}\label{Kequation}
\tilde{\cal K}\, {\bf H}\cdot\nabla_v\mathcal{D}+ {\bf H}\,
\tilde{\cal K}\cdot\nabla_u\mathcal{D}&=& \lambda\sum_{i=1}^r
\,e^{\alpha_i\cdot(B^T u+B v)/2}\,[E_{\alpha_i},\tilde{\cal K}]-
\frac{1}{\lambda}\,E_{-\alpha_1}\,\tilde{\cal
K} \,e^{\alpha_1\cdot B(u-v)/2} \nn\\
&&-\frac{1}{\lambda} \sum_{i=2}^r\left(E_{-\alpha_i}\,\tilde{\cal
K}- \tilde{\cal
K}\,E_{-\alpha_{i-1}}\right)e^{\alpha_i\cdot B(u-v)/2}.
\end{eqnarray}
Bearing in mind the form of the defect potential
\eqref{DefectPotentialConformal+free}, and assuming the
following perturbation solution
for $\tilde{\cal K}$
\begin{equation}\label{kExpansion}
\tilde{\cal K}=I+\sum_{i=1}^{\infty}\,\frac{k_{i}}{\lambda^{i}},
\end{equation}
the terms on either side of the expression \eqref{Kequation} must match
at each order in $\lambda$. It is straightforward to see
that this happens for terms of order $\lambda$ and $\lambda^0$
provided $k_i=\sigma\sum^r_{i=1}\,E_{-\alpha_i}$.
However, the terms at order $1/\lambda$
are  trickier to analyze. They give
\begin{eqnarray}\label{Oder1/lambda}
\sigma\,\sum_{i=1}^r(E_{\alpha_i} {\bf H}\cdot\nabla_v {\cal B}+{\bf
H} E_{\alpha_i}\cdot\nabla_u \mathcal{D})&=& \sum_{i=1}^r
\,e^{\alpha_i\cdot(B^T u+B v)/2}\,[E_{\alpha_i},k_2]-E_{-\alpha_1}
\,e^{\alpha_1\cdot B(u-v)/2} \nn\\
&&-\sum_{i=2}^r\left(E_{-\alpha_i}
- E_{-\alpha_{i-1}}\right)\,e^{\alpha_i\cdot B(u-v)/2}.
\end{eqnarray}
Making use of the defect potential once more \eqref{DefectPotentialConformal+free},
and of the explicit expression \eqref{Bmatrix0}
for the matrix $B$, it is possible to compare separately the terms in $1/\lambda$
proportional to $\exp{(\alpha_i\cdot B(u-v)/2)}$ and
$\exp{(\alpha_i\cdot(B^T u+B v)/2)}$.
The former lead to
\begin{equation}\label{Oder1/lambda1half}
-\sum^r_{ij=1}\,\frac{1}{2}(\alpha_j\cdot B\alpha_i)\,
e^{\alpha_i\cdot B(u-v)/2}=-\sum_{i=2}^r\left(E_{-\alpha_i}
- E_{-\alpha_{i-1}}\right)\,e^{\alpha_1\cdot B(u-v)/2}-E_{-\alpha_i}
\,e^{\alpha_1\cdot B(u-v)/2},
\end{equation}
where the Lie algebra commutation relations \eqref{LieAlgebra} have been used.
The expression \eqref{Oder1/lambda1half} is clearly an identity.
The remaining terms of \eqref{Oder1/lambda}, which are proportional to
$\exp{(\alpha_i\cdot B(u-v)/2)}$, lead to an expression for
$k_2$\footnote{See \cite{Bowcock04} for details of a similar calculation}.
On the other hand, $k_2\equiv 0$ when evaluated in an $(r+1)$ dimensional
representation for which
\begin{equation}
(E_{\alpha_i})_{ab}=\delta_{ai}\,\delta_{bi-1}, \qquad a,b=1,\dots,(r+1).
\end{equation}
Therefore, in this particular representation a complete expression for the
element $\tilde{\cal K}$ is
\begin{equation}
\tilde{\cal K}=I+\frac{\sigma}{\lambda}\sum_{i=1}^{\infty}\,E_{-\alpha_i}.
\end{equation}

\p
The existence of a Lax pair representation strongly suggests that the
system described in this section is integrable. Note, it is worth emphasizing
that a carefully chosen collection of defects arranged along
 the $x$-axis is able to link an $a_r$ conformal
Toda field theory with $r$ free massless fields. Or, since the $a_1$ Toda field model
is so related to a massless field, the $a_r$ Toda model can be
decomposed into a collection
of $r$ Liouville models instead, or indeed to a mixture of $p$ Liouville models
and $q$ massless free fields with $(p+q)=r$.

\p Note, in the discussion above to solve expression \eqref{Kequation} it was
supposed that $\tilde{\cal K}$ was
an expansion in inverse powers of the
spectral parameter $\lambda$. In \cite{Bowcock04} it was pointed out
that $\tilde{\cal K}$ could also be regarded as having  an expansion
in positive powers
of $\lambda$. In those circumstances a slightly different - yet still consistent -
 relationship between the matrices $A$, $B$ and the
form of the defect potential $\mathcal{D}$, was found. In the present
case, by looking at \eqref{Kequation}, it should be noted that such a
possibility is not allowed.
In fact, to be able to obtain an alternative solution it would be necessary
to start with a different expression to \eqref{TheLaxPair}
for the Lax pair describing an $a_{r-1}$ Toda theory together with a
free massless field. The Lax pair representation
\eqref{TheLaxPair} may be replaced by:
\begin{eqnarray}\label{TheLaxPairAlternative}
a_t&=& \frac{1}{2}\left[\partial_x v \cdot
{\bf H}+ \sum_{i=1}^{r-1}\left(\lambda\,
E_{\alpha_i}-\frac{1}{\lambda}\,
E_{-\alpha_i}\right)e^{\alpha_i\cdot v/{2}}\right]-
\frac{1}{\lambda} \,E_{-\alpha_r}\, e^{\alpha_r\cdot v/2}, \nonumber\\
a_x&=&\frac{1}{2}\left[\partial_t v \cdot
{\bf H}+ \sum_{i=1}^{r-1}\left(\lambda\,
E_{\alpha_i}+\frac{1}{\lambda}\,
E_{-\alpha_i}\right)e^{\alpha_i\cdot v/{2}}\right]+
\frac{1}{\lambda} \,E_{-\alpha_r}\, e^{\alpha_r\cdot v/2},
\end{eqnarray}
which leads - via the zero curvature condition - to the same equations of motion
\eqref{EquationOfMotion}. Proceeding in a similar manner as before, and using the
same matrix $B$ \eqref{Bmatrix0}, the analogue of expression \eqref{Kequation} is
\begin{eqnarray}\label{KequationAlternative}
\tilde{\cal K}\, {\bf H}\cdot\nabla_v\mathcal{D}+ {\bf H}\,
\tilde{\cal K}\cdot\nabla_u\mathcal{D}&=& \lambda \sum_{i=2}^r
\left(E_{\alpha_i}\,\tilde{\cal
K}-\tilde{\cal
K}\,E_{\alpha_{i-1}}\right)e^{-\alpha_i\cdot B(u-v)/2}+ \lambda
\,E_{\alpha_1}\,\tilde{\cal
K}\,e^{-\alpha_1\cdot B(u-v)/2}\nn\\
&&\hskip 10pt
-\frac{1}{\lambda}\sum_{i=1}^r
\,e^{-\alpha_i\cdot(B^T u+B v)/2}\,[E_{-\alpha_i},\tilde{\cal K}],
\end{eqnarray}
which can be solved using an expansion in positive powers of $\lambda$ for
the element $\tilde{\cal K}$. It should be mentioned that to achieve this
a slightly different relationship between the matrices $A$ and $B$
has been used, namely
\begin{equation}\label{MKMatrixRelationAlternative}
B=-(1+A), \qquad B^T=(-1+A)=(-2-B).
\end{equation}
Expression \eqref{MKMatrixRelationAlternative} can be obtained by the
total momentum conservation analysis of section (\ref{MomentumConservation})
by looking at
the first expression in \eqref{MomentumConservationConstraints1set}.
In fact, it can be rewritten in an alternative way with respect to
\eqref{MKMatrixRelation} as
\begin{equation}
(-1+A^T)(-1+A)=BB^T,
\end{equation}
from which \eqref{MKMatrixRelationAlternative} follows.

\section{Classical $a_r$ affine Toda models with a defect}
\label{ClassicalDefect}

\p In this section attention will be focussed on the affine Toda model related to
the Lie algebra $a_r$.  To summarize briefly, the model is described by the following
Lagrangian density
\begin{equation}\label{TodaDefect}
\mathcal{L}_{D}=\theta(-x)\mathcal{L}_u +\theta(x)\mathcal{L}_v-
\delta(x)\left(\frac{1}{2}u_t\cdot
A u +\frac{1}{2}v_t \cdot A v+u_t\cdot B v+\mathcal{D}(u, v )\right).
\end{equation}
The bulk Lagrangian densities $\mathcal{L}_u$ and $\mathcal{L}_v$ are given by
\eqref{TodaL} with all integers $n_j$ equal to one, and
 $(\alpha_j \cdot \alpha_j)=2$. The matrix $B=(1-A)$, which is given by the
formula \eqref{Bmatrix0}, and satisfies the following:
\begin{equation}\label{constraintsonB}
\alpha_k\cdot B\alpha_j=\left\{%
\begin{array}{ll}
    \phantom{-}2 & \hbox{$k=j$,} \\
    -2 & \hbox{$k=j+1$,} \\
    \phantom{-}0 & \hbox{otherwise,} \\
\end{array}%
\right.\qquad j=0,\dots, r,\qquad
\alpha_{r+1}=\alpha_0.
\end{equation}
Finally, the defect potential $\mathcal{D}$ is given in \eqref{DefectPotential0}
where $\sigma$ is the defect parameter.
Setting $r=1$ the Lagrangian \eqref{TodaDefect} describes the
sinh-Gordon model with a purely transmitting defect, first investigated
from this point of view in in \cite{bczlandau}.

\p The $a_r$ affine Toda model with fields and coupling constant ${\beta}$
restricted to be real describes, after quantization, $r$ interacting scalars,
also known as fundamental Toda
particles, whose classical mass parameters are given by
\begin{equation}\label{Todaparticlemasses}
m_a=2\,m\sin\left(\frac{\pi a }h\right), \quad a=1,2\dots,r,
\end{equation}
where $h=(r+1)$ is the Coxeter number of the algebra. On the other
hand, if the fields are permitted to be complex the model possesses
classical `soliton' solutions \cite{Hollowood92}. Conventionally,
in the description of the complex affine Toda field theory the
coupling constant $\beta$ is
replaced with $i \beta$. It is then easy in \eqref{TodaDefect}
to switch from the real affine Toda model
for the Lie algebra $a_r$ to the complex one.
In the bulk  soliton solutions interpolate between
constant zero energy field configurations as $x$ runs from $-\infty$ to
$\infty$. The constant solutions are given by $v=2\pi \lambda/\beta$, where
$\lambda$ belongs to the weight lattice of the Lie algebra $a_r$.
Each of them is characterized by a topological charge, which is defined as follows
\begin{equation}\label{topologicalcharges}
Q=\frac{\beta}{2\pi}\int^{\infty}_{-\infty}dx\, u_x=
\frac{\beta}{2\pi}\left[\phi(\infty,t) -\phi(-\infty,t)\right],
\end{equation}
and lies in the weight lattice of the algebra.
Explicitly, solutions of this type have the form
\begin{equation}\label{TodaSolitonu}
u_{a}=-\frac{1}{i\beta}\sum^{r}_{j=0}\alpha_{j}\ln \left(1+
E_a\,\omega^{j}_a\right), \quad E_a=e^{a_a x-b_a t +\xi_a},\quad
\omega_a=e^{2\pi i a/h},\quad a=1,\dots,r
\end{equation}
where $(a_a, b_a)=m_{a}\, (\cosh{\theta},\sinh{\theta})$, $\theta$
is the soliton rapidity and  $\xi_a$ is
a complex parameter, which, though almost arbitrary, must be chosen so
that there are no singularities in the solutions as the real
coordinates $x$ and $t$ vary.
Despite the solutions \eqref{TodaSolitonu} being complex, Hollowood
\cite{Hollowood92}
showed that their total energy and momentum is actually real and
their masses, at rest,
are given by
\begin{equation}\label{singlesolitonmasses}
M_a=\frac{2\,h\,m_a}{\beta^2}, \quad a=1,2\dots,r,
\end{equation}
where $m_a$ are the mass parameters of the real scalar theory
\eqref{Todaparticlemasses}.

\p For each $a=1,\dots, r$ there are several solitons
whose topological
charges lie in the set of weights of the fundamental $a^{th}$
representation of $a_r$ \cite{McGhee94}. However, apart from the two extreme
cases, $a=1$ and $a=r$, not every weight belonging to one of the other
representations corresponds to the topological charge of a stationary soliton.
The number $\tilde{n}_a$ of possible charges for the representation
with label $a$ is exactly equal
to $h$ divided by the greatest common divisor of $a$ and $h$.
By shifting the parameter $\xi_a$ by $2\pi a/h$
the soliton solution \eqref{TodaSolitonu} changes its topological
charge, since such a shift operates a cycle permutation of the roots
($\alpha_j\rightarrow \alpha_{(j-1)}$). Such a permutation is
equivalent to the application of the Coxeter element
\begin{equation}
t(\alpha_j)=s_rs_{n-1}\dots s_2s_1(\alpha_j),\qquad s_i(\alpha_j)=
\alpha_j-(\alpha_j\cdot \alpha_i)\alpha_i.
\end{equation}
Therefore, the relevant weights are orbits
of the Coxeter element.

\p When a defect is introduced, some of the properties previously described
will change. For instance, constant field configurations,
which are solutions of both the equations of motion and the defect conditions
that follow from the Lagrangian \eqref{TodaL},
are given by $(u,\, v)=(2\pi \lambda_a/\beta,\, 2\pi \lambda_b/\beta)$,
where the label $a$ and $b$ refer to the specific fundamental representations to
which the weights $\lambda_a$ and $\lambda_b$ belong (up to translations by roots,
since energy and momentum
are invariant under translations of the fields by elements of the root lattice).
Their energy and
momentum is now different from zero and equal to (and for convenience, $\sigma=e^{-\eta}$)
\begin{equation}\label{defectenergies}
({\cal E}_{a,b},\, {\cal
P}_{a,b})=-\frac{2hm}{\beta^2}\left[\cosh\left(\eta-\frac{2 (a-b)\pi
i}{h}\right),\ -\sinh\left(\eta-\frac{2 (a-b)\pi i}{h}\right)\right],\
\ a,b=1,\dots, r.
\end{equation}
Notice that when the two weights describing the static
configurations of the fields $u$ and $v$ belong to the same
representation
the energy and momentum will be
real.\footnote{The parameter $\eta$ is chosen to be real.}
Notice also that the topological charges carried by a defect constitute a much
larger set than the number of possibilities for stationary solitons themselves.

\p Another interesting change introduced by the defect is represented by the
behaviour of a soliton solution which travels through a defect.
By convention, a soliton \eqref{TodaSolitonu} with positive rapidity
will travel from
the left to the right along the $x$-axis and at some time it will meet the defect
located at $x=0$. The soliton $v$ emerging on the right will be
similar to $u$, but  delayed. It is  described by,
\begin{equation}\label{TodaSolitonv}
v_{a}=-\frac{1}{i\beta}\sum^{r}_{j=0}\alpha_{j}\ln \left(1+
z_a\,E_a\,\omega^{j}_a\right),
\end{equation}
where $z_a$ represents the delay of the soliton travelling through
the defect and which, by making use of the defect conditions, is found to be
\begin{equation}\label{ClassicalDelay}
z_a=\left(\frac{e^{\,-(\theta-\eta)}-i\,e^{i\pi a/h}}
{e^{\,-(\theta-\eta)}-i\,e^{-i\pi a/h}}\right), \qquad \sigma=e^{-\eta}.
\end{equation}
This delay is generally  complex with  exceptions being
self-conjugate solitons, corresponding to $a=h/2$ (with $r$ odd),
for which the delay is real. The expression \eqref{ClassicalDelay}
has a complex simple pole
at $\theta=\eta+i\left(\frac{\pi a}{h}-\frac{\pi}{2}\right)$.
This means that a soliton with real rapidity can be absorbed by
a defect only if it lies in the self-conjugate representation.
This fact was first noticed in \cite{bczlandau} in the context
of the sine-Gordon model. In \cite{Corrigan07}, by examining
the argument of the phase of the delay \eqref{ClassicalDelay},
it was noticed that the defect might induce a phase shift
in the soliton that effects a change in the topological
charge of the soliton itself, at least provided the shift lies
in a suitable range. It was found that the phase shift
can be at most equal to $2\pi a/h$ for $a=1,\dots, (h-1)/2$
($r$ even), or $a-1,\dots, r/2-1$ ($r$ odd). While it is
$-2\pi a/h$ for the corresponding anti-solitons $(h-a)$.
This quantity should be compared with the quantity separating
two different topological charge sectors, which is $2\pi/\tilde{n}_a$.
This suggests that a soliton in the first representation or an
anti-soliton in the corresponding last representation might
convert, at most, to one of the adjacent solitons/anti-solitons
within its multiplet as it passes the defect. However, the
scope for  jumping to configuration other than adjacent solitons
increases as the representation investigated moves towards
representations associated with more central spots of the Dynkin diagram.

\section{A transmission matrix for the $a_r$ affine Toda field theories}

\p In \cite{Corrigan07} the transmission matrix for the $a_2$ affine Toda
model was thoroughly investigated.  A complete classification
 of the infinite dimensional solutions of the triangular equation - subject only to a
few reasonable assumptions - were obtained. Among them, it was
possible to select solutions relevant for the defect problem, and to
complete them with a suitably chosen (though not unique) overall scalar factor
fixing their zero-pole
structure in a minimal way. In this section, the aim is to extend those
results to the whole
$a_r$ affine Toda series. In \cite{Corrigan07} the
different behaviour of solitons and anti-solitons travelling through a
defect was noted. In particular, it was always possible to find a solution for which
one group (the $a=1$ solitons, for example) seemed to match the strict selection
rule mentioned above at the end of section (\ref{ClassicalDefect}), which
concerned the restricted possibilities for a soliton to change its topological charge,
while the other group ($a=2$) did not. In a sense this was surprising albeit entirely
consistent with the requirements of the bootstrap. On the other hand, some differences
between solitons and anti-solitons should be expected because of the lack of parity
or time-reversal invariance of the Lagrangian describing the defect conditions.
It will be seen that this different
behaviour between solitons and anti-solitons is found in all $a_r$
affine Toda models, at least for solitons and anti-solitons in the
first ($a=1$) and last ($a=r$) representations, respectively.

\p The starting point is the set of `triangle relations' that relate the
elements of the transmission matrix $T$ to the elements of the bulk
scattering matrix $S$ \cite{Delf94}. They are:
\begin{equation}\label{STT}
S{_{kl}^{mn}}(\Theta)\,T{_{n\alpha}^{t\beta}}(\theta_1)\,
T{_{m\beta}^{s\gamma}}(\theta_2)=T{_{l\alpha}^{n\beta}}(\theta_2)\,
T{_{k\beta}^{m\gamma}}(\theta_1)\,S{_{mn}^{st}}(\Theta),
\end{equation}
where $\Theta=(\theta_1-\theta_2)$. Note the presence of two types of labels in the
transmission matrix elements. The roman
labels are a finite set of positive integers $1,2,\dots,d$ labelling the soliton states
within a representation of dimension $d$, while the greek labels
represent vectors in  the weight lattice of the Lie algebra $a_r$
(it is expected that a stable, basic defect will be labelled by the root lattice).

\p The $S$-matrices describing the scattering of solitons in the
$a_r$ affine Toda field theory were conjectured some time ago by
Hollowood \cite{Hollowood93}.
Hollowood's proposal makes use of Jimbo's $R$-matrices \cite{Jimbo89},
which are trigonometric solutions of the
Yang-Baxter equation (YBE) associated with the quantum group $U_q(a_r)$.
According to the proposal, the solitons of the model
lie in (and fill up) the $r$ different multiplets corresponding for generic $q$
to the $r$
fundamental representations of the algebra $U_q(a_r)$.
The number of states in each multiplet coincides with the number of weights in
the corresponding representation. For example, the $S$-matrix $S^{ab}(\Theta)$
describes the scattering of two solitons with rapidities
$\theta_1$ and $\theta_2$, lying in the multiplets $a$ and $b$,
respectively. Hence, it is an interwining map on the two representation spaces
$V_a$ and $V_b$
\begin{equation}\label{TriangularEquation}
S^{ab}(\Theta): V_a\otimes V_b \longrightarrow V_b \otimes
V_a,\qquad S^{ab}(\Theta)=\rho^{ab}(\Theta)\,R^{ab}(\Theta)
\end{equation}
where $R^{ab}$ is Jimbo's $R$-matrix and $\rho^{ab}$ is a
scalar function determined by the requirements of `unitarity',
crossing symmetry, analyticity and other consistency requirements (such as bootstrap
relations), which a scattering matrix ought to satisfy
\cite{Hollowood93}.

\p However, in practice it is enough to know explicitly the $S^{11}$-matrix -
also known as the fundamental scattering matrix
- describing the scattering of the solitons in the first representation,
since all the other scattering matrices can be obtained from it on
applying a bootstrap procedure.
The representation space $V_1$ of the first multiplet has
dimension $h$ and its states are labeled by  weights
representation, which can be written conveniently as follows:
\begin{equation}\label{Weights1}
l_{j}^1\equiv
l_j=\sum_{l=1}^r\,\frac{(h-l)}{h}\,\alpha_l-\sum_{l=1}^{j-1}\,\alpha_l,
\quad j=1,\dots,h.
\end{equation}
Abbreviating $S^{11}\equiv S$, the non-zero elements of $S$ are given by:
\begin{eqnarray}\label{S11elements}
S^{jj}_{jj}\,(\Theta)&=&\rho(\Theta)\,
\left(q\,X-q^{-1}\,X^{-1}\right),
\nn\\
&&\nn\\
S^{kj}_{jk}\,(\Theta)&=&\rho(\Theta)\,
\left(X-X^{-1}\right), \quad k\neq j, \nn\\\nn\\
S^{jk}_{jk}\,(\Theta)&=&\rho(\Theta)\,\left(q-q^{-1}\right)\left\{%
\begin{array}{ll}
X^{(1-2|l|/h)}\left.\right|_{\,l=j-k<0}\\
\\
X^{-(1-2|l|/h)}\left.\right|_{\,l=j-k>0}\\
\end{array}%
\right.
\end{eqnarray}
with
\begin{equation}
X=\frac{x_1}{x_2},\qquad x_j=e^{h\gamma \theta_j/2},\quad j=1,2\qquad
q=-e^{-i\pi\gamma}, \qquad
\gamma=\frac{4\pi}{\beta^2}-1.
\end{equation}
(Note, it is assumed $0<\beta^2<4\pi$ so that $0<\gamma<\infty$.)

\p The scalar function $\rho$ is given by the following expression
\begin{eqnarray}\label{rhofunction}
\rho(\Theta)&=&
\,\,\frac{\Gamma(1+h\gamma i\,\Theta/2\pi)
\Gamma(1-h\gamma i\,\Theta/2\pi-\gamma)}{2\pi i}\,\,
\frac{\sinh(\Theta/2+i\pi/h)}{\sinh(\Theta/2-i\pi/h)}\nn\\
&&\qquad\qquad \times\ \prod_{k=1}^\infty\,
\frac{F_k(\Theta)\,F_k(2\pi i/h-\Theta)}{F_k(2\pi i
/h+\Theta)\, F_k(2\pi i -\Theta)},
\end{eqnarray}
where
$$
    F_k(\Theta)=\frac{
    \Gamma(1+h\gamma i\,\Theta/2\pi+hk\gamma)}
    {\Gamma(h\gamma i\,\Theta/2\pi+(hk+1)\gamma)
    }.
$$
Equipped with the $S$-matrix, it is possible, in principle, to solve the triangle
equations \eqref{TriangularEquation} to obtain
an expression for the transmission matrix $T^1$ (which will be
denoted by $T$) for  solitons lying in the first
representation.
As a consequence of  topological
charge conservation, the elements of the transmission matrix
will have the following form
\begin{equation}\label{Telements}
T{^{n\beta}_{i\alpha}}(\theta)=
t{_{i\alpha}^{n}}(\theta)\;\delta^{\beta-l_i+l_n}_{\alpha}\qquad
i,n=1,\dots,h,
\end{equation}
where $l_i,l_n$ are the weights \eqref{Weights1}
and $\alpha$ and $\beta$ lie in the root lattice. However, when this expression
is inserted into equation \eqref{TriangularEquation}, one rapidly discovers
there are many different solutions.
 Amongst these there will be the transmission matrix that describes the scattering
of a soliton by a defect, itself characterized classically by the choice of the
matrix $B$ \eqref{Bmatrix0}.  Making use of both the experience acquired in this
kind of calculation and the results already obtained for the $a_2$ affine Toda model,
it reasonable to claim that the non zero elements
of the appropriate solution - up to an undetermined scalar function $g(\theta)$ - are:
$$
T^{i\beta}_{i\alpha}(\theta)=g(\theta)\,Q^{\alpha \cdot
l_i}\,\delta^{\beta}_{\alpha},\quad
T^{(i-1)\beta}_{i\alpha}(\theta)=g(\theta)\,(t^{1/h}\,x^{2/h})\,
\delta^{\beta-l_i+l_{i-1}}_{\alpha},
\quad i=1,\dots, h,
\quad (i-1)=0\equiv h,
$$
where
$t$ is a constant parameter and $Q=-e^{i\pi\gamma}$ depends on the coupling constant
appearing in the classical Lagrangian density.
Setting $t=e^{-h\gamma\Delta}$ and $\hat{x}=e^{\gamma( \theta-\Delta)}$, the solution
above can be rewritten in the following neater form
\begin{equation}\label{Tmatrix1}
T^{i\beta}_{i\alpha}(\theta)=g(\theta)\,Q^{\alpha \cdot
l_i}\,\delta^{\beta}_{\alpha},\qquad
T^{(i-1)\beta}_{i\alpha}(\theta)=g(\theta)\,\hat{x}\,
\delta^{\beta-l_i+l_{i-1}}_{\alpha},
\quad i=1,\dots, h,\quad (i-1)=0\equiv h.
\end{equation}
It should be pointed out that suitably designed
unitary transformations - of the same type as those used  in
\cite{Corrigan07} - have been used to reduce the number of free constants appearing
in the solution to just the one essential
parameter $\Delta$. Notice that solution \eqref{Tmatrix1} provides a good match
with the classical situation because of the presence of zeros in
expected positions, meaning that a soliton might convert to only one
of its adjacent solitons, thereby respecting the classical selection
rules mentioned earlier. Further calculations using the bootstrap 
- which will not appear in this 
article - suggest that this agreement between 
the classical and the quantum situation 
with regard to selection rules holds also for the other `soliton' representations $a$ 
with $a=2,3,\dots, (h-1)/2$ ($r$ even) or $a=2,3,\dots, (h/2-1)$ ($r$ odd), but not
for the rest, regarded as `anti-soliton' representations. 

\p The transmission matrices describing the interaction between a defect and the
solitons lying in any of the other representations of the algebra $a_{r}$ could be
computed by applying a bootstrap procedure, which will be described in the
next section. Such a procedure together with an additional constraint is
also used to obtain the overall factor $g(\theta)$. The argument goes as
follows. First of all, the extra constraint is provided by the crossing
relation that the solution \eqref{Tmatrix1} must satisfy,  given by
\begin{equation}\label{Crossing}
T^{(h-a)}{_{n\alpha}^{i\beta}}(\theta)=\tilde{T}^a{_{i\alpha}^{n\beta}}
(i\pi-\theta) \quad a=1,\dots,r
\end{equation}
where the matrix $\tilde{T}^a$ describes
the interaction between the defect and a soliton within the $a$
representation travelling from the
right to the left. In fact, since parity is violated explicitly
in the description of the
defect, the matrix $\tilde{T}^a$ is expected to
differ from the matrix $T^a$ describing solitons
travelling from left to right. Obviously, the matrix $\tilde{T}^a$ itself
satisfies a set of triangular equations albeit with a different, though related,
$S$-matrix.  These equations differ in some details from
\eqref{STT}, and therefore $\tilde{T}^a$ is not amongst its solutions. Note,
however, that matrices $T^a$ and $\tilde{T}^a$ must be related to each other,
and it is natural to suppose the following:
\begin{equation}\label{Unitarity}
T^a{_{a\alpha}^{b\beta}}(\theta)\,\tilde{T}^a{_{b\beta}^{c\gamma}}(-\theta)
=\delta^{c}_{a}\delta^{\gamma}_{\alpha}.
\end{equation}
Notice that for the sine-Gordon model, which is the only affine Toda field theory
in the $a_r$ series to be unitary, the expression \eqref{Unitarity}
is equivalent to the unitarity condition since in that case
$\tilde{T}(-\theta)\equiv(T(\theta))^{\dagger}$. Thus, by computing the
inverse of solution \eqref{Tmatrix1}, it is possible to obtain the
transmission matrix - again up to a multiplicative factor - for
the solitons within  the representation $r$, which are in fact  anti-solitons
with respect to the solitons in the first representation. The elements of this
matrix read
\begin{equation}\label{Tmatrixr}
T^r{^{(i+j)\beta}_{i\alpha}}(\theta)=\frac{\hat{x}^{j}\,Q^{-\alpha \cdot
k_{(i+j)}}}{g(\theta-i\pi)(1-\hat{x}^h\,Q^{-1})}\,\delta^{\beta+l_i-l_{i+j}}_{\alpha},
\end{equation}
with
$$i=1,\dots, h\qquad j=0,\dots, (h-1)\qquad k_{(i+j)}=l_i+l_{i+1}+\dots +l_{i+j},$$
and where it must be borne in mind that $(i+j)$ is evaluated mod($h$).
It should be remarked that the weights in the representation $r$ are
$-l_i$, with $l_i$ given by \eqref{Weights1}. Notice that this time,
the solution \eqref{Tmatrixr} does not possess the expected zeros
corresponding to the classical selection rule. Each anti-soliton
may convert into any of the anti-soliton within the same representation,
though the classically allowed transmission remains the most probable.

\p Comparing the solution \eqref{Tmatrixr}, obtained by
applying the crossing relation, with the solution for the same anti-solitons
that will be computed in the next section, it is possible to constrain
the overall function $g(\theta)$ and find an explicit
expression for it.

\section{Bootstrap procedure and the overall factor of the transmission matrix}

\p Consider $D_{\alpha}$ to be a formal operator representating
the defect. Then, it is natural to describe the
interaction between a defect and a soliton within the
representation  $a$ as follows
\begin{equation}\label{DefectOperator}
A^a_{i}(\theta)D_{\alpha}=T^a{^{j\beta}_{i\alpha}}(\theta)D_{\beta}A^a_{j}(\theta),
\end{equation}
where $A^a_i$ is set of operators representing the soliton state $i$ in the
representation $a$. The total number of states in the representation $a$ is
$h!/(a!(h-a)!)$, and, in principle, by making use of the $h$ states within the first
representation, all other states can be built. Hence,
expression \eqref{DefectOperator} allows to construct all transmission
matrices simply relying on the $T^{1}\equiv T$-matrix \eqref{Tmatrix1}.
The construction of the soliton states can be elucidated using an iterative process.
Consider the states $l_k^2$ in the second representation. Since each weight
is $l_i^2=l_j+l_k$ where $l_j$, $l_k$ are the weights \eqref{Weights1} with
$j\neq k$. The corresponding state is given, schematically, by
\begin{equation}\label{StatesIn2}
A^2_{i\{jk\}}(\theta)\equiv {^{11}}c_i{^{jk}}\,A_j(\theta-i\pi/h)\,
A_k(\theta+i\pi/h)+{^{11}}c_i{^{kj}}\,A_k(\theta-i\pi/h)\,
A_j(\theta+i\pi/h),
\end{equation}
where $\Theta=i 2\pi/h$ is the location of the simple pole in the scattering
matrix $S^{11}\equiv S$ corresponding to a soliton in the second representation.
The constants ${^{11}}c_i{^{jk}}$ and ${^{11}}c_i{^{kj}}$ are the couplings,
whose ratio - effectively the only data needed - can be calculated using the
scattering matrix $S$.

\p The next representation is the third one, and to
construct its states more care must be taken. The argument goes
as follows. The weights associated with each state can be written as
$l_i^3=l_j+l_k+l_m$ with $j\neq k\neq m$, that is making use only of the
weight in the first representation. Formally, such a state can be formulated
as follows
\begin{eqnarray}\label{StatesIn3S12}
A^3_{i\{jkm\}}(\theta)&\equiv& {^{12}}c_i{^{jp}}\,A_j(\theta-i2\pi/h)\,
A^2_{p\{km\}}(\theta+i\pi/h)+{^{12}}c_i{^{kq}}\,A_k(\theta-i2\pi/h)\,
A^2_{q\{jm\}}(\theta+i\pi/h)\nn\\
&&+{^{12}}c_i{^{mt}}\,A_m(\theta-i2\pi/h)\,
A^2_{t\{jk\}}(\theta+i\pi/h),
\end{eqnarray}
where the coupling ratios can be calculated using the scattering matrix $S^{12}$.
The pole in this matrix corresponding to a soliton in the third representation
is located at $\Theta=i3\pi/h$. Note that an equivalent formulation could have
been provided by using the matrix $S^{21}$. The relevant pole is still located
at $\Theta=i3\pi/h$, but the expression for a soliton state in the third
representation would have been
\begin{eqnarray}\label{StatesIn3S21}
A^3_{i\{jkm\}}(\theta)&\equiv& {^{21}}c_i{^{pj}}\,
A^2_{p\{km\}}(\theta-i\pi/h)\,A_j(\theta+i2\pi/h)+{^{21}}c_i{^{qk}}\,A^2_{q\{jm\}}
(\theta-i\pi/h)\,A_k(\theta+i2\pi/h)\nn\\
&&+{^{21}}c_i{^{tm}}\,A^2_{t\{jk\}}(\theta-i\pi/h)\,A_m(\theta+i2\pi/h).
\end{eqnarray}
In fact, given three soliton states
$A_{j}(\theta_1)\,A_{k}(\theta_2)\,A_{m}(\theta_3)$,
expression \eqref{StatesIn3S12} describes the case in
which first the solitons described by $A_{k}(\theta_2)$ and $A_{m}(\theta_3)$
combine together to form a soliton $A_{p}^2(\theta')$ and
subsequently,  $A_{j}(\theta_1)$ and $A_{p}^2(\theta')$
form the soliton $A_{q}^3(\theta'')$. On the other hand, expression \eqref{StatesIn3S21}
corresponds to a situation where the solitons represented by $A_{j}(\theta_1)$
and $A_{k}(\theta_2)$  combine first to give a soliton in the second
representation, and so on. 

\p Similarly, and with even more care, it is possible to construct all soliton
states on recognizing that the pole corresponding to a soliton in the $c$
representation is located at $\Theta={i(a+b)\pi/h}$ in the scattering matrix
$S^{ab}$ with $c=a+b$.

\p Applying \eqref{DefectOperator} to the soliton states in the $r$ representation,
it is possible to find
\begin{equation}\label{Tmatrixr'}
T^r{^{(i+j)\beta}_{i\alpha}}(\theta)=f_r(\theta)\,\hat{x}^{j}\,Q^{-\alpha \cdot
k_{(i+j)}}\,\delta^{\beta+l_i-l_{i+j}}_{\alpha},
\end{equation}
where
\begin{equation}
f_r(\theta)=\prod^{r/2-1}_{a=0}\,g(\theta-i(2a+1)\pi/h)\,g(\theta+i(2a+1)\pi/h),
\end{equation}
if $r$ is even, and
\begin{equation}
f_r(\theta)=g(\theta)\,\prod^{(r-1)/2}_{a=1}\,g(\theta-i2a\pi/h)\,g(\theta+i2a\pi/h),
\end{equation}
if $r$ is odd. Note that the latter formula holds for $r\neq 1$, since the bootstrap
cannot be applied in the sine-Gordon situation. In that case, one has
simply $f_r(\theta)=g(\theta)$.
Finally, the solution \eqref{Tmatrixr'} may be compared with the solution
\eqref{Tmatrixr}, to provide a constraint for the scalar function $g(\theta)$.
 It reads
\begin{equation}
g(\theta+i2\pi)=g(\theta)\frac{(1+\hat{x}^h\,(-Q)^{r})}{(1+\hat{x}^h\,(-Q)^{r+2})}.
\end{equation}
for which a minimal solution  - for all $a_r$ affine Toda models - is:
\begin{equation}\label{gFunction}
g(\theta)=\frac{\hat{g}(\theta)\,\hat{x}^{-1/2}}{2\pi}
\end{equation}
\p with
\begin{equation}
\hat{g}(\theta)=\Gamma[1/2+(r/2)\gamma-z]\prod_{k=1}^{\infty}
\frac{\Gamma[1/2+(hk+r/2)\gamma-z]\,
\Gamma[1/2+(hk-1-r/2)\gamma+z]} {\Gamma[1/2+(hk-r/2)\gamma-z]\,
\Gamma[1/2+(hk-r/2)\gamma+z]},
\end{equation}
where $$\hat{x}=e^{\gamma( \theta-\Delta)},\qquad z=
\frac{ih\gamma(\theta-\Delta)}{2\pi}.$$
The technique adopted in this section can be extended to all
representations, and in principal all transmission matrices $T^{a}$ with
$a=1,\dots,r$ can be found.
For the overall scalar
function $g^a(\theta)$ a compact formula reads
\begin{equation}\label{gaFunction}
g^a(\theta)=\frac{\hat{g}^a(\theta)\,\hat{x}^{-a/2}}{2\pi}\qquad a=1,\dots,r
\end{equation}
\p with
\begin{eqnarray}
\hat{g}^a(\theta)&=&\Gamma[1/2+(h-a)\gamma/2-z]\times\nonumber \\
&&\hskip 15pt \prod_{k=1}^{\infty}
\frac{\Gamma[1/2+(hk+(h-a)/2)\gamma-z]\,
\Gamma[1/2+(hk-(h+a)/2)\gamma+z]} {\Gamma[1/2+(hk-(h-a)/2)\gamma-z]\,
\Gamma[1/2+(hk-(h-a)/2)\gamma+z]}.
\end{eqnarray}
The $\Gamma$-function outside the product contains an interesting
complex pole, which is located at
\begin{equation}\label{Pole}
\theta_a =\Delta -i\vartheta_a,\qquad \vartheta_a=\frac{\pi(h-a)}{h} +
\frac{\pi}{h\gamma}\qquad a=1,\dots,r.
\end{equation}
Comparing this pole with the pole appearing in the classical delay
\eqref{ClassicalDelay},
it is possible to relate the defect parameter $\sigma$ to the
complex parameter $\Delta$. Given that in the classical limit $1/\gamma\rightarrow 0$,
the identification of the two poles \eqref{Pole}
and \eqref{ClassicalDelay} requires 
$$\Delta=\eta+\frac{i\pi}{2},\qquad \sigma=e^{-\eta}.$$
The complex energy associated with this pole is
\begin{equation}\label{PoleEnergy}
E_a=M_a\cosh\theta_a=M_a\cosh\eta\, \sin\vartheta_a
+i\,M_a\sinh\eta \,
\cos\vartheta_a,
\end{equation}
where $M_a$ is the mass of a soliton in the representation $a$ given by
\eqref{singlesolitonmasses}. Provided \eqref{PoleEnergy} enjoys a positive real
part and a negative imaginary part, that is
\begin{equation}\label{PoleConstraint}
\pi/2\leq\vartheta_a<\pi,
\end{equation}
the pole \eqref{Pole} corresponds to an unstable bound state.

\p Consider first a soliton lying in a representation labelled by $a\leq(h/2-1)$ or $a\leq(h-1)/2$,
depending whether $h$ is even or odd. Then, bearing in mind that $1/\gamma$ is
always a positive - or, in the classical limit, zero - quantity,
condition \eqref{PoleConstraint}
is satisfied provided $1/\gamma<(h/2-1)$ or
$1/\gamma<(h-1)/2$, respectively. Note that in the classical limit
($1/\gamma \rightarrow 0$) the energy \eqref{PoleEnergy} is typically complex and
appears to correspond to the energy of an unstable bound state, which could be
identified in the classical version of the model as one of the defects whose
energy is given by \eqref{defectenergies} (taking $u=0,\, v=2\pi\lambda_a/\beta$).
In fact, only if $h$ is even and  $a=h/2$ does $\vartheta_a\rightarrow \pi/2$
in the classical limit, meaning the imaginary part of \eqref{PoleEnergy}
disappears leaving a
real part equal to the energy of a soliton $a$, which moves with
rapidity $\eta$. This situation corresponds to the classical
possibility for a self-conjugate soliton to be infinitely
delayed by the defect. Finally, if the soliton lies in a
representation, $a\geq(h/2+1)$ or
$a\geq(h+1)/2$,\, (meaning it is an `anti-soliton' according
to the convention used so far), again, depending whether $h$ is even or
odd,  an unstable bound state appears  within a range $(a-h/2)\leq 1/\gamma<a$
of the coupling
that does not include a neighbourhood of the classical limit. In other words,
these quantum unstable states would be  disconnected from any phenomenon
occurring in the classical models. That the different representations behave
differently in this context appears to compound the difficulties in comparing the
quantum theory of these models with the classical theory; not only do real states
go missing but unstable states appear unexpectedly. Perhaps these phenomena are related.

\newpage

 \noindent{\bf Acknowledgements} \vskip .25cm
\p  This
article grew out of a review
talk given by one of us (EC) at a meeting dedicated to Alexei Zamolodchikov, 
his life  and work, at
the J.-V. Poncelet French-Russian Mathematical Laboratory  in Moscow. We both knew
Aliosha, as a friend and  colleague, and we shall miss his words of advice and 
encouragement.

\p We also wish to express our gratitude to the UK Engineering and Physical Sciences
Research Council for its support under the grant EP/F026498/1 and to the 
Galileo Galilei Institute, Florence, for its hospitality in September-October 2008; 
one of us (EC)
also wishes to thank Ryu Sasaki and the Yukawa Institute for Theoretical Physics for
its hospitality during July 2008.

\end{document}